\documentclass{article}
\usepackage{graphicx}
\usepackage{booktabs}
\usepackage{hyperref}
\usepackage{xurl}

\title{SoLiD26: A First Principles Solid-Liquid Interface Dataset for Machine-learned Interatomic Potentials}

\author{
Jonas Busk$^{1,2,*}$
\and Emil J. P. Frost$^{1,2}$\\
\and Yogeshwaran Krishnan$^1$\\
\and Henrik H. Kristoffersen$^1$\\
\and August E. G. Mikkelsen$^1$\\
\and Xueping Qin$^1$\\
\and Xin Yang$^1$\\
\and Heine A. Hansen$^{1,2}$\\
\and Arghya Bhowmik$^{1,2}$\\
\and Tejs Vegge$^{1,2,*}$
}
\date{}

\begin{document}

\maketitle

\begin{center}
  {\footnotesize $^1$Department of Energy Conversion and Storage,\\Technical University of Denmark, Kongens Lyngby, Denmark.\\ \vspace{.5em}
  $^2$Pioneer Center for Accelerating P2X Materials Discovery (CAPeX),\\Technical University of Denmark, Kongens Lyngby, Denmark.\\ \vspace{.5em}
  $^*$Corresponding authors: Jonas Busk (jbusk@dtu.dk), Tejs Vegge (teve@dtu.dk).}
\end{center}

\begin{abstract}

Machine-learned interatomic potentials (MLIPs) enable simulations of larger atomistic systems over longer timescales than traditional ab initio methods but require large, diverse training data. Existing datasets lack comprehensive sampling of solid-liquid interfaces, which exhibit complex phenomena such as solvent layering and surface reconstruction critical for energy and catalytic materials. Here, we present SoLiD26, a curated solid-liquid interface dataset of 15.4 million first-principles structures with up to 576 atoms and 15 chemical elements for training and evaluating MLIPs. The structures were compiled from density functional theory (DFT) calculations, with most configurations originating from ab initio molecular dynamics (AIMD) simulations. SoLiD26 includes aqueous coinage metal interfaces, electrode–electrolyte systems, and selected bulk reference structures, calculated with VASP using the PBE functional and D3 dispersion corrections. The application of SoLiD26 for training and evaluating MLIPs is demonstrated with a suite of MACE models on a simple training, validation and test split. This dataset provides a valuable resource for advancing MLIP development targeted at modelling solid-liquid interfaces in electrochemical and catalytic systems.
\end{abstract}


\section*{Background \& Summary}

\begin{figure}[b]
\centering 
\includegraphics[width=0.14\linewidth]{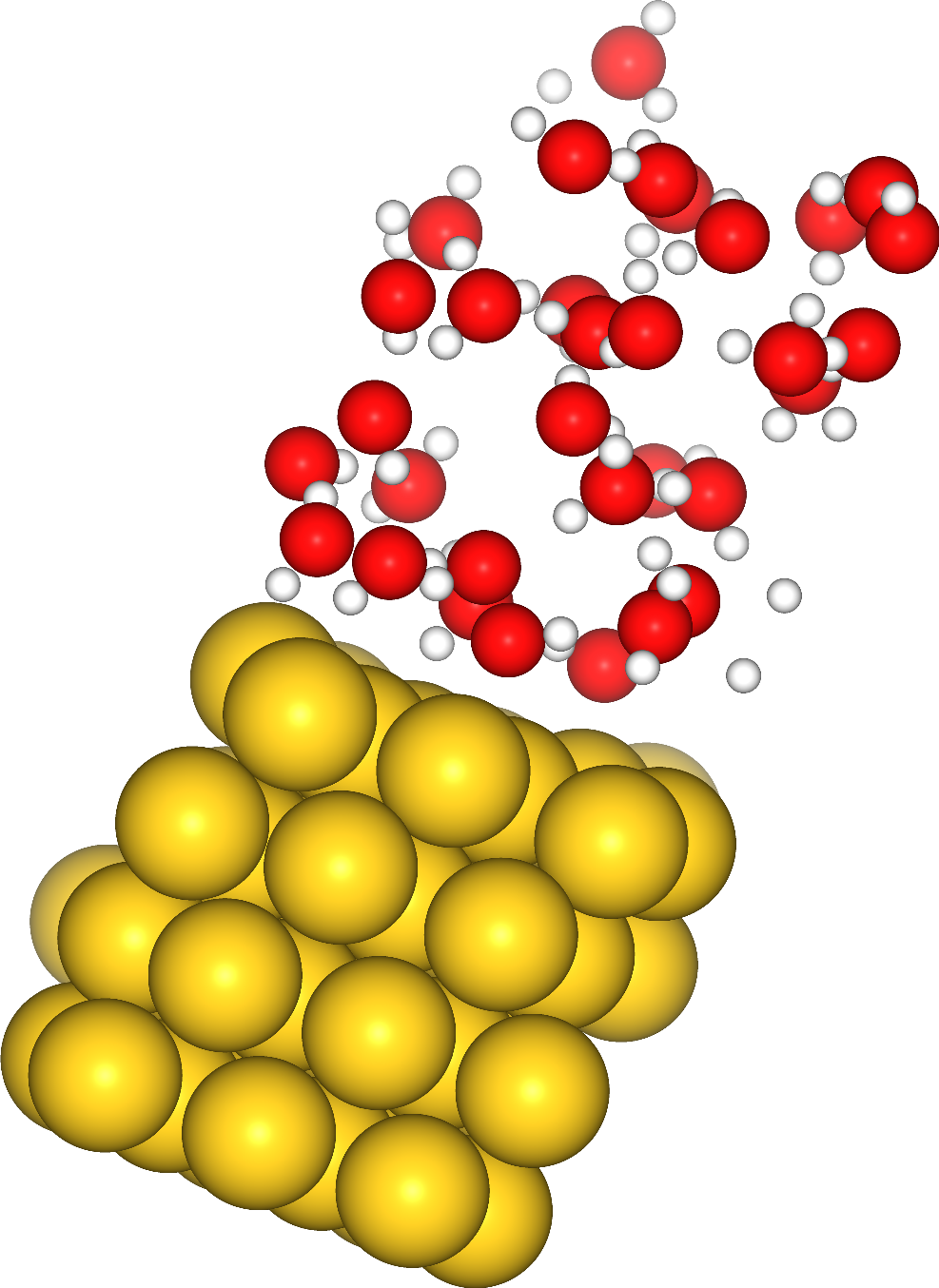}
\hspace{1em}
\includegraphics[width=0.14\linewidth]{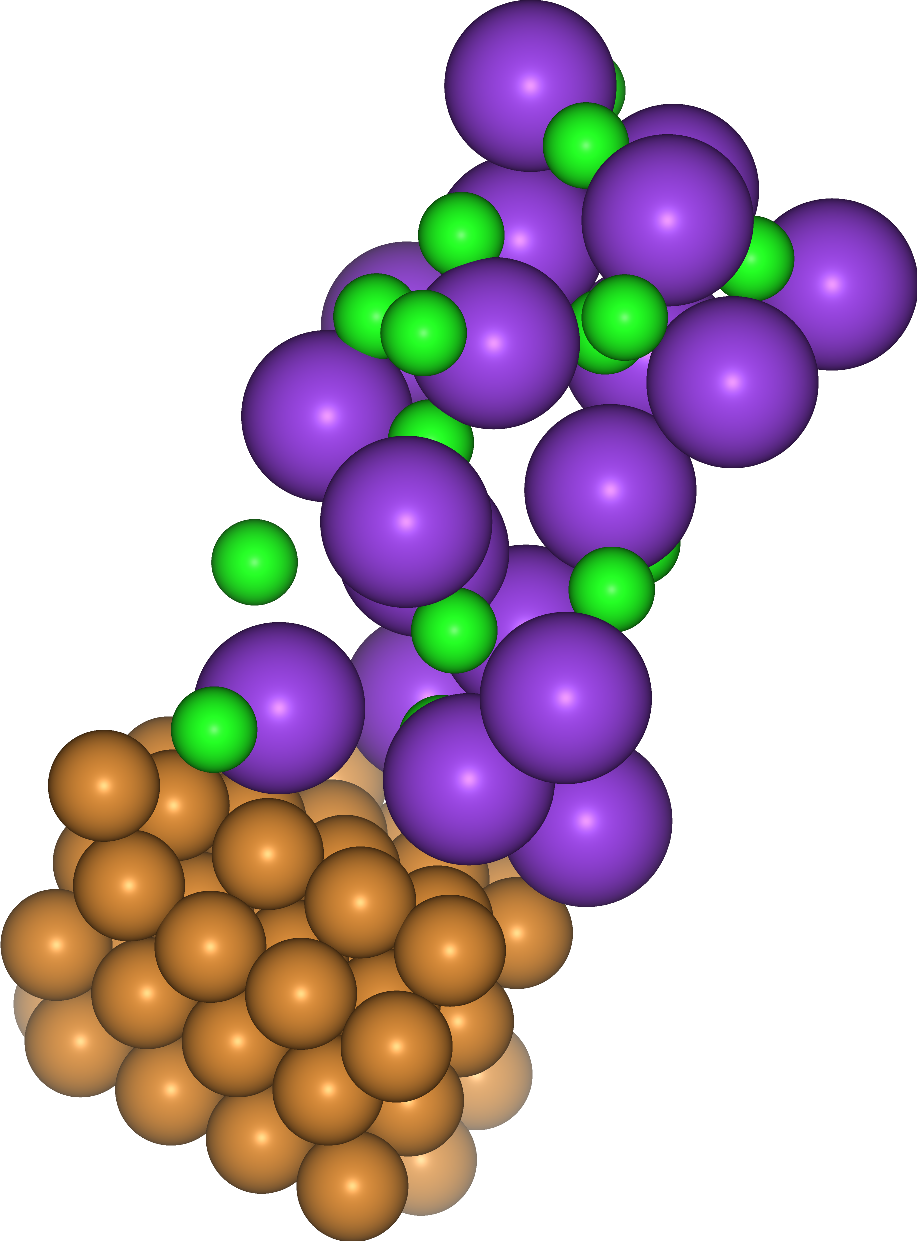}
\hspace{1em}
\includegraphics[width=0.14\linewidth]{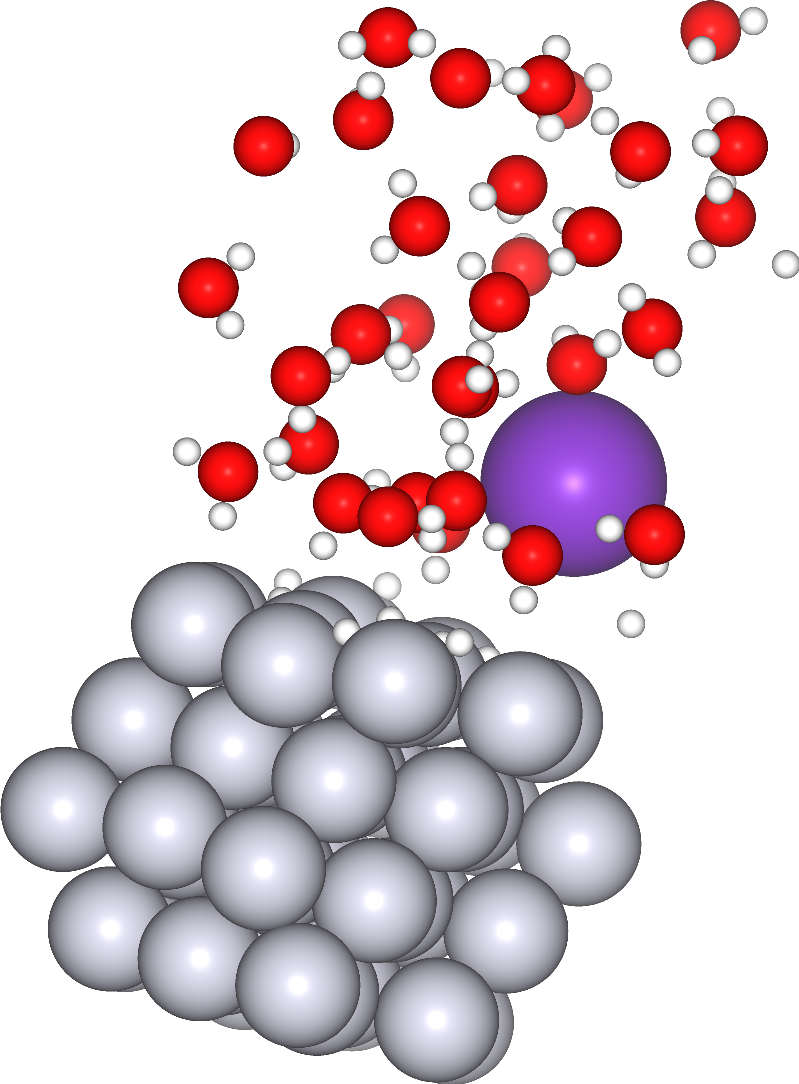}
\hspace{1em}
\includegraphics[width=0.14\linewidth]{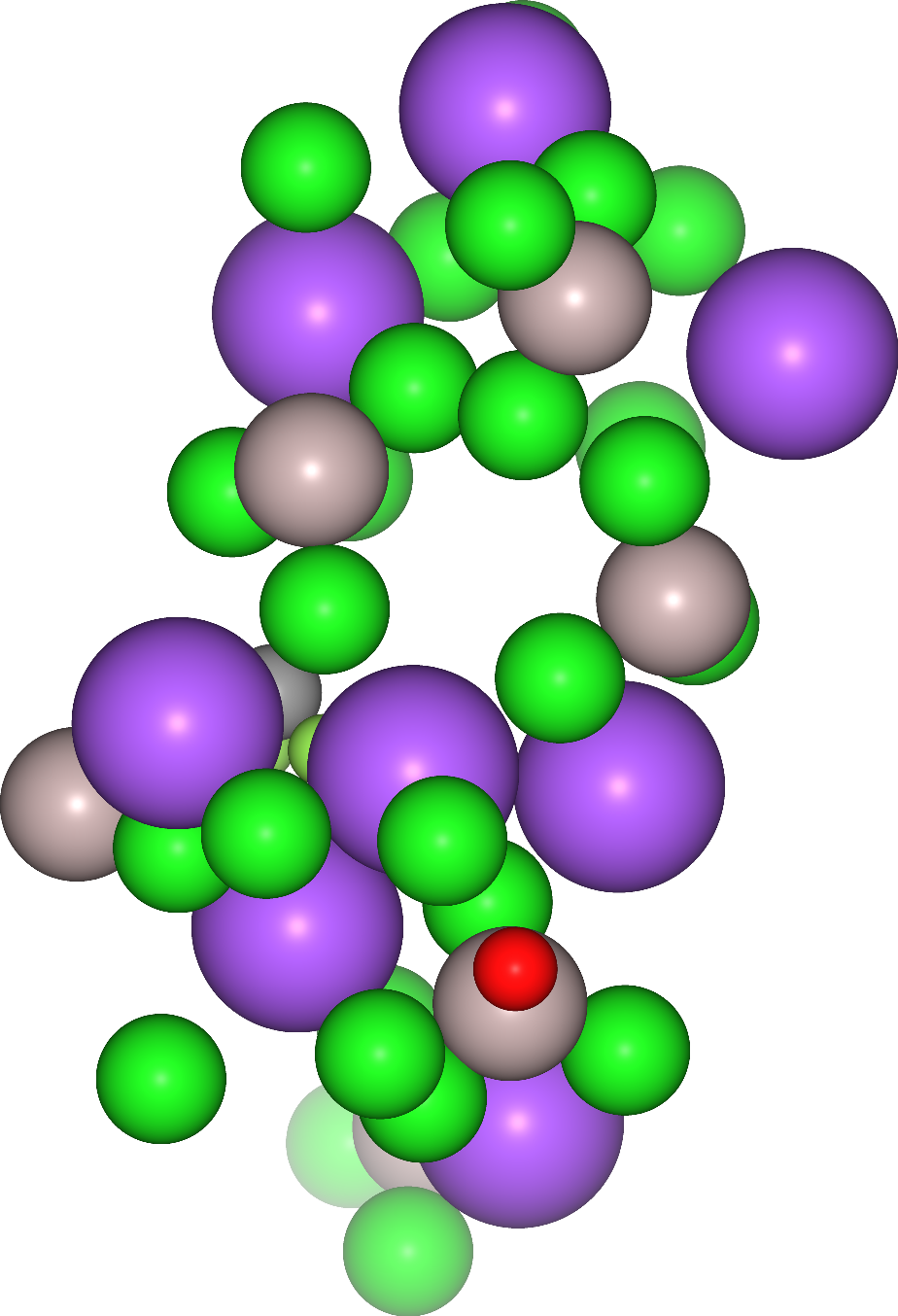}
\hspace{1em}
\includegraphics[width=0.14\linewidth]{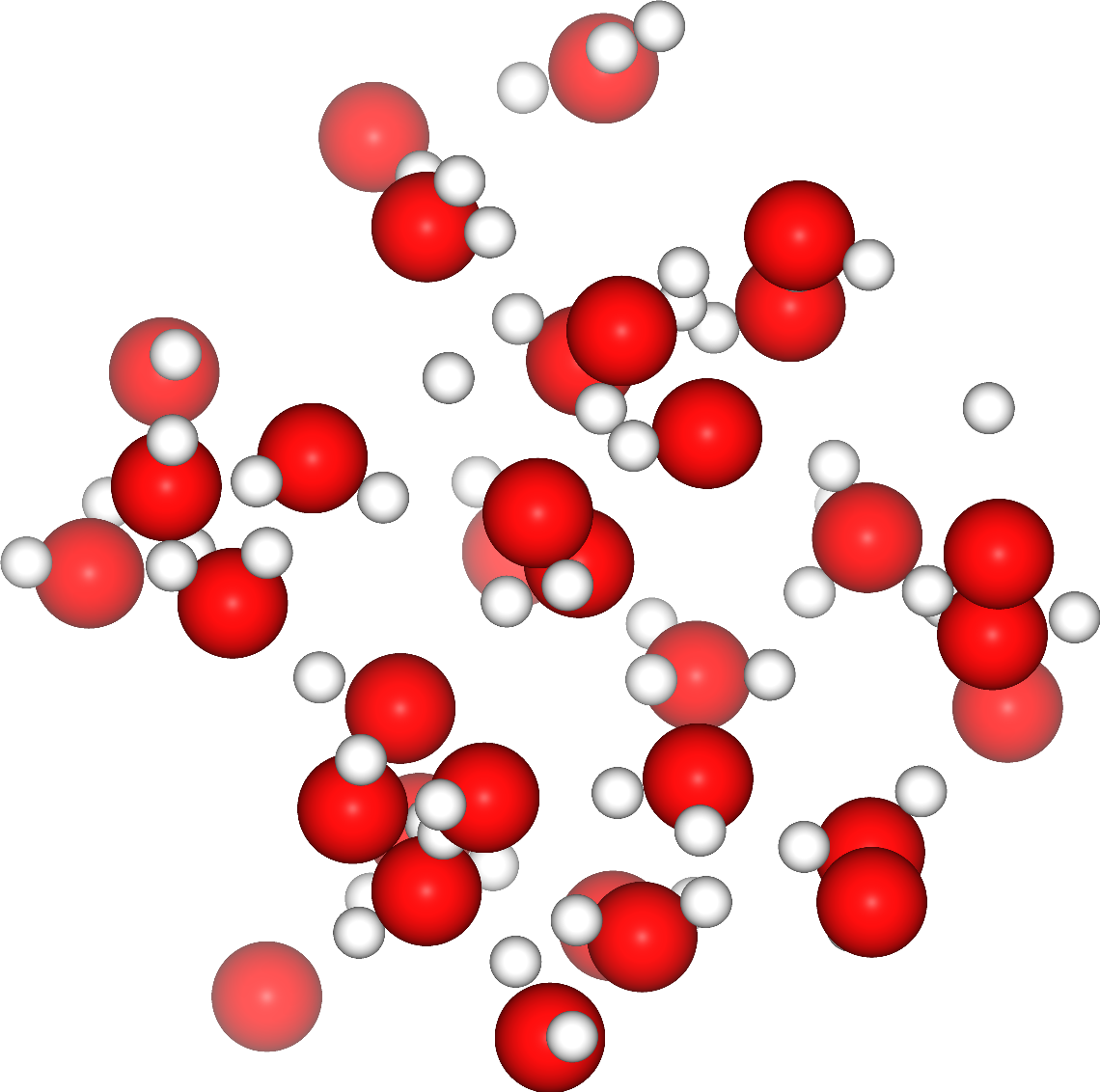}
\caption{Example structures from the SoLiD26 dataset. The dataset contains 15,379,823 structures with up to 576 atoms spanning 15 different elements and consists primarily of solid-liquid interfaces and some bulk reference structures.}
\label{fig:structures}
\end{figure}

Machine-learned interatomic potentials (MLIPs) can help accelerate materials discovery by enabling simulation of larger atomistic systems over longer time scales than is currently feasible with state-of-the-art ab initio methods such as density functional theory (DFT).
Recent advances in atomistic machine learning foundation models for materials have demonstrated high accuracy across a range of chemical systems~\cite{mace-mp, pet-mad}.
Achieving this generalisation, however, requires large and diverse training datasets that span a wide variety of chemistries.

Existing datasets for atomistic machine learning predominantly describe isolated molecules, bulk crystals, relaxed structures or high-temperature trajectories of relatively small systems~\cite{chgnet, schreiner2022, nandi2023, omat24, matpes}. While these datasets provide an important foundation for broad pre-training, they do not systematically sample the coupled response of a liquid and a solid surface. Explicit solid–liquid interfaces exhibit phenomena such as solvent layering, surface reconstruction, ion adsorption, solvent-mediated bond formation and breaking, and fluctuations across multiple length and time scales. These configurations are also considerably larger than those found in many commonly used molecular and bulk datasets, making direct first principles sampling computationally demanding. Only a few datasets, such as the Open Catalysis 2025 (OC25)~\cite{oc25}, focus specifically on interfaces and such datasets therefore complement general-purpose training datasets and provides targeted data for fine-tuning and evaluating MLIPs in electrochemical and catalytic environments.

Understanding and predicting the spatio-temporal evolution of solid-liquid interfaces in advanced energy materials, such as batteries and catalysts, remains a significant challenge due to the kinetic barriers and long-range electrostatic interactions that govern rate-limiting reactions and processes~\cite{bhowmik2019}.
In these systems, the interface is exposed to electric fields and chemical reactions, resulting in properties that are distinct from those of the bulk phases. Therefore, bespoke training data targeting solid-liquid interfaces is necessary to enable MLIPs to accurately capture this domain.

In this work, we consolidate first principles calculations generated in several studies of aqueous metal interfaces, electrochemical reactions and electrode–electrolyte systems at DTU Energy~\cite{august2021, august2022, henrik2018, henrik2020, xueping2021, xueping2023, xueping2024a, xueping2024b, xin2023, xin2024, brandes2024} into a large, coherent dataset. 
The curated release contains 15,379,823 structures with up to 576 atoms, spanning the elements H, Li, C, N, O, F, Na, Al, Cl, K, Ca, Cu, Cs, Pt, and Au (Figure~\ref{fig:structures}). In addition to interfacial configurations, the dataset includes selected liquid and solid reference structures from the source studies. The release is intended for training, fine-tuning and benchmarking MLIPs, particularly in regimes requiring simultaneous sampling of solid, liquid, ionic, and reactive configurations.
The utility of the dataset is validated by training MLIPs and evaluating their predictive accuracy on held out data.
The following sections describe the dataset preparation, validation and availability in detail.

\section*{Methods}

The SoLiD26 dataset was prepared by collecting and curating data from several research projects at DTU Energy that investigated solid-liquid interfaces for energy and catalysis applications using computational methods~\cite{august2021, august2022, henrik2018, henrik2020, xueping2021, xueping2023, xueping2024a, xueping2024b, xin2023, xin2024, brandes2024}.
The majority of the data was generated through ab initio molecular dynamics (AIMD) simulations, providing significant variation of energy and forces, essential for learning the potential energy surface (PES) with MLIPs.
The data ingestion and preparation pipeline was composed of a series of Python scripts utilising the ASE package~\cite{ase} and involved the following overall steps:
First, atomic structures with energy and forces calculated with VASP using the PBE functional and D3 dispersion correction method~\cite{vasp} were identified and collected in an initial dataset.
The initial dataset was then filtered to ensure high data quality and consistency.
Finally, a clean version of the dataset was prepared for publication.

The data filtering process was designed to prepare the dataset for machine learning applications by excluding outliers through the following steps:
Calculations with energy cutoff (encut) values other than 350, 400 or 450 were excluded.
Structures with a maximum force greater than 10 eV/Å were excluded.
Duplicate structures (for computational efficiency defined as structures with identical chemical formula, energy, maximum force, and centre of mass) were excluded.
Structures containing elements represented in less than one percent of the dataset were excluded.
Additional outliers were excluded by identifying structures with high prediction errors on forces in a 5-fold cross-validation machine learning experiment, under the assumption that such errors are primarily an indication of chemical outliers or unconverged DFT calculations.
For this machine learning-based filtering, we employed PaiNN models~\cite{painn}, which are relatively fast and computationally inexpensive to train and evaluate while providing sufficient accuracy for outlier detection. The dataset was randomly split into 5 equal-sized parts, with one part held out from the training of each of the 5 models. The models were then evaluated on the respective held-out parts and structures with a high prediction error on forces were excluded from the dataset.

\section*{Data Records}

The dataset described in this manuscript is available from the Materials Cloud Archive~\cite{solid26mca}. The dataset is provided in ASE~\cite{ase} format as a collection of 16 sequentially numbered \texttt{.aselmdb} files. 
Each structure in the dataset is stored as an \texttt{ase.Atoms} object (\url{https://docs.ase-lib.org/ase/atoms.html}), representing a collection of atoms described by atomic numbers, positions, cell, periodic boundary conditions, potential energy and atomic forces.
The data files are compressed with \texttt{xz} and the uncompressed files are approximately 100 GB in total. 
The release includes MD5 checksums for each file and a JSON file with the indices of the training, validation and test splits used for the technical validation described in this work.
Please refer to the Usage Notes section for details about how to read the data files.

\section*{Data Overview} 

\begin{figure}[h]
\centering 
\includegraphics[width=0.95\linewidth]{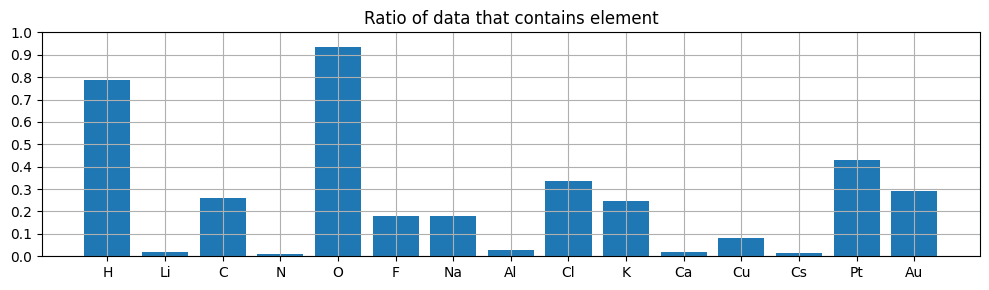}
\caption{The elements represented in SoLiD26. The figure shows the proportion of all the structures in the dataset that include each element. For example, 78.6~\% of all the structures in the dataset contain H.}
\label{fig:elements}
\end{figure}

The dataset contains a total of 15,379,823 atomic structures with up to 576 atoms and an average of 126.3 atoms, including the elements: H, Li, C, N, O, F, Na, Al, Cl, K, Ca, Cu, Cs, Pt, and Au (Figure~\ref{fig:elements}).
The structures primarily represent solid-liquid interfaces with periodic boundary conditions, but the dataset also includes some bulk structures, such as water and gold.
The potential energy and atomic forces for each structure were calculated using VASP with the PBE functional and D3 dispersion correction~\cite{vasp}.

\section*{Technical Validation}

To demonstrate and validate that SoLiD26 is useful for training and fine-tuning MLIPs in the domain of solid-liquid interfaces, we have evaluated a suite of MACE models~\cite{mace}: The MACE-MATPES-PBE-0 pre-trained model~\cite{mace-mp, matpes} with added D3 dispersion correction~\cite{grimme2010, grimme2011, takamoto2022} through the ASE calculator included in the MACE codebase (\url{https://github.com/acesuit/mace}), a MACE-MATPES-PBE-0 model fine-tuned on SoLiD26 (MACE-SoLiD26-ft), and a MACE model trained from scratch on SoLiD26 (MACE-SoLiD26) with similar hyperparameters as the MACE-MATPES-PBE-0 model, which, according to the release notes (\url{https://github.com/ACEsuit/mace-foundations}), is based on the MACE-OMAT-0 medium pre-trained model~\cite{mace-mp, omat24, batatia2025}.
The MACE-MATPES-PBE-0 model was fine-tuned by simply loading the pre-trained model and continued training on SoLiD26 with a low learning rate of 0.0005 until convergence, also known as naive fine-tuning~\cite{tompa2026}.
The training and fine-tuning were both done in two stages by first training with a loss weight of 1 on the energy and 10 on the forces followed by additional training with a loss weight of 10 on the energy and 1 on the forces to reduce the final energy error~\cite{kovacs2023}.
All the training and evaluation was performed on NVIDIA H100 and H200 GPUs. 

The dataset was split into training, validation and test sets as follows: All structures with more than 200 atoms (272,889) were designated as the test set to ensure that the test data is distinct from the training data and to assess whether the resulting models can generalize to larger structures. However, this simple approach only tests the models on a subset of the chemistries and only a subset of the elements (H, C, O, K, Pt, Au) is represented in the test set. Further generalisation experiments are left for future work.
From the remaining data, 100,000 structures were randomly selected for the validation set to track training progress, and the rest (15,006,934 structures) were used as the training set. 

The results of evaluating the three different MACE models on the held-out test dataset are presented in Table 1.
The pre-trained MACE-MATPES-PBE-0 model with added D3 dispersion correction serves as a solid baseline, although it was not trained specifically on solid-liquid interface data. Both the MACE-SoLiD26-ft and MACE-SoLiD26 models show significant improvements compared to the baseline, as expected for models trained specifically on in-domain data. Substantial improvements are seen in the energies, but in particular on the forces, where the MAE and RMSE are reduced to approximately one third compared to MACE-MATPES-PBE-0. This strongly indicates that SoLiD26 can be used to improve models in the domain of solid-liquid interfaces.
It should be noted that for this technical validation of the dataset, the model hyperparameters were not systematically optimised, so further performance improvements are expected in future work.

\begin{table}[]
    \footnotesize
    \centering
    \begin{tabular}{lrrrrrr}
        \toprule
        & \multicolumn{2}{c}{\bf Energy (eV)} & \multicolumn{2}{c}{\bf Energy PA (eV)} & \multicolumn{2}{c}{\bf Forces (eV/Å)} \\
        \cmidrule(lr){2-3}\cmidrule(l){4-5}\cmidrule(l){6-7}
        \bf Model & MAE & RMSE & MAE & RMSE & MAE & RMSE \\
        \midrule
        MACE-MATPES-PBE-0 & 1.3341 & 2.0506 & 0.0046 & 0.0065 & 0.0501 & 0.0696 \\
        MACE-SoLiD26-ft   & 0.8665 & 1.5807 & 0.0027 & 0.0041 & 0.0118 & 0.0253 \\ 
        MACE-SoLiD26      & 0.8866 & 1.6790 & 0.0027 & 0.0043 & 0.0118 & 0.0236 \\ 
        \bottomrule
    \end{tabular}
    \caption{Results of evaluating MACE models on SoLiD26. The table shows mean absolute error (MAE) and root mean squared error (RMSE) on energy, energy per atom and forces for each model. The pre-trained (MACE-MATPES-PBE-0) and fine-tuned (MACE-SoLiD26-ft) models are based on the MACE-MATPES-PBE-0 foundation model and the MACE model trained from scratch (MACE-SoLiD26) uses similar hyperparameters.}
    \label{tab:placeholder}
\end{table}

\section*{Usage Notes} 

The dataset files can be read using Python by installing the \texttt{ase} (\url{https://pypi.org/project/ase/}) and \texttt{ase-db-backends} (\url{https://pypi.org/project/ase-db-backends/}) packages.
With both packages installed, the standard \texttt{ase.db.connect()} method will recognize the \texttt{.aselmdb} file format and the data can be used as described in the ASE database documentation (\url{https://docs.ase-lib.org/ase/db/db}).
Some atomistic machine learning packages already support the \texttt{.aselmdb} file format and have dataset classes that can be used to load the dataset, such as \texttt{fairchem-core} (\url{https://fair-chem.github.io/}) and \texttt{mace-torch} (\url{https://github.com/acesuit/mace}).

Users of SoLiD26 should be aware that the dataset contains structures (\texttt{ase.Atoms} objects) that have specified constraints, such as fixed atoms, which were defined in the original calculations. This means that using the standard \texttt{atoms.get\_forces()} method in ASE may inadvertently set some forces to zero. Instead use \texttt{atoms.get\_forces(apply\_constraint=False)}, or get the forces directly from the underlying data structure, to access the raw, unmodified forces, as using the altered forces in a machine learning context can lead to erroneous results.

\section*{Data Availability}

SoLiD26 is available from the Materials Cloud Archive and can be accessed via the following URL: \url{https://doi.org/10.24435/materialscloud:yp-vw}.

\section*{Code Availability}

The code used to prepare SoLiD26 is available from the CAPeX GitHub repository at \url{https://github.com/team-capex/SoLiD26}.
The repository also includes the scripts used to train the MACE-SoLiD26-ft and MACE-SoLiD26 models as well as the trained models evaluated in this work.

\bibliographystyle{unsrturl}
\bibliography{references}

\section*{Author Contributions}

JB: Methodology, Software, Formal analysis, Data Curation, Writing - Original Draft, Visualization.
EF: Methodology, Formal analysis, Data Curation, Writing - Original Draft.
YK, HK, AM, XQ, XY: Investigation.
HH, AB: Writing - Review \& Editing, Supervision.
TV: Conceptualization, Writing - Original Draft, Supervision, Project administration, Funding acquisition.

\section*{Competing Interests}

No competing interests.

\section*{Acknowledgements} 

The authors would like to thank Clara Hollenbeck and Jakob Cetti Hansen for carrying out preliminary data exploration.

\section*{Funding}

The authors acknowledge support from the Danish National Research Foundation: Pioneer Center for Accelerating P2X Materials Discovery (CAPeX), DNRF grant number P3, and from the Novo Nordisk Foundation Data Science Research Infrastructure 2022 Grant: A high-performance computing infrastructure for data-driven research on sustainable energy materials, Grant no. NNF22OC0078009, and Grants for access to the Gefion AI Supercomputer April 2025, Grant no. NNF25OC0105158.

\end{document}